\documentclass[twocolumn]{aastex62}
\submitjournal{ApJL}

\usepackage{amsthm, amsmath, amssymb}
\usepackage{latexsym,graphicx,rotating,amsmath, epsfig, natbib, graphbox}

\newcommand{\del}{\nabla}

\renewcommand{\vec}{\boldsymbol}
\newcommand{\pomega}{\varpi}

\newcommand{\scrS}{\mathcal{S}}

\newcommand{\Ra}{\mathrm{Ra}}
\newcommand{\Ek}{\mathrm{Ek}}
\renewcommand{\Pr}{\mathrm{Pr}}
\newcommand{\Pm}{\mathrm{Pm}}
\newcommand{\RoCsq}{\mathrm{Ro}_\mathrm{C}^2}
\newcommand{\RoC}{\mathrm{Ro}_\mathrm{C}}

\newcommand{\dedalus}{\href{http://dedalus-project.org/}{Dedalus}}

\begin{document}

\title{Single-hemisphere dynamos in M-dwarf stars}
\correspondingauthor{Benjamin  Brown}
\email{bpbrown@colorado.edu}

\author[0000-0001-8935-219X]{Benjamin P.\ Brown}
\affiliation{Department Astrophysical and Planetary Sciences \& LASP,
University of Colorado,
Boulder, CO 80309, USA}

\author[0000-0001-8531-6570]{Jeffrey S.\ Oishi}
\affiliation{Department of Physics and Astronomy,
Bates College,
Lewiston, ME 04240, USA}

\author[0000-0002-8902-5030]{Geoffrey M.\ Vasil}
\affiliation{School of Mathematics and Statistics, University of Sydney,
Sydney, Sydney, NSW 2006, AUS}

\author[0000-0002-7635-9728]{Daniel Lecoanet}
\affiliation{Princeton Center for Theoretical Science \& Department of Astrophysical Sciences,
  Princeton University,
Princeton, NJ 08544, USA}

\author[0000-0003-4761-4766]{Keaton J.\ Burns}
\affiliation{Department of Mathematics, Massachusetts Institute of Technology, Cambridge, MA 02139, USA}
\affiliation{Center for Computational Astrophysics, Flatiron Institute,
New York, NY 10010, USA}

\begin{abstract}
M-dwarf stars below a certain mass are convective from their cores to their photospheres.
These fully convective objects are extremely numerous, very magnetically active, and the likely hosts of many exoplanets.
Here we study, for the first time, dynamo action in simulations of stratified, rotating fully convective M-dwarf stars.
Importantly, we use new techniques to capture the correct full ball geometry down to the center of the star.
We find surprising dynamo states in these systems, with the global-scale mean fields confined strongly to a single hemisphere, in contrast to prior stellar dynamo solutions.
These hemispheric-dynamo stars are likely to have profoundly different interactions with their surroundings, with important implications for exoplanet habitability and stellar spindown.
\end{abstract}

\section{Introduction} \label{sec:intro}

Tiny M-dwarfs are the most numerous stars in the universe and one of the most promising locations for finding habitable planets.
In particular, they have many Earth-sized habitable planets, and the nearest habitable exoplanet, when found, is likely to orbit an M-dwarf \citep[e.g.,][]{Dressing_Charbonneau_2015}.
M-dwarfs are intrinsically dim, and observations of them are hard, but our data set is growing rapidly and now includes long-duration campaigns of slowly-rotating systems \citep{Newton_et_al_2018}.
The habitability of potential exo-Earths around other stars is at the mercy of the magnetic environment. M-dwarfs, in particular, display multi-kilogauss surface fields.
Photometric monitoring infers incredibly large starspots \citep[e.g.,][]{Newton_et_al_2016} or many, many small spots \citep[e.g.,][]{Jackson_Jeffries_2012}, and the combination of strong fields and complex topologies leads exceptionally high levels of magnetic activity \citep[e.g.,][]{West_et_al_2015}.
Mega-flares larger than any ever seen on the Sun \citep[e.g.,][]{Schmidt_et_al_2019} happen often enough to threaten the ability of planets to support life, or possibly even to retain atmospheres \citep[e.g.,][]{Shkolnik_et_al_2014}.

The interiors of type-M4 stars or later (masses of $\sim 0.35M_\odot$ or less) are convectively unstable from their nuclear burning cores to the photosphere.
Despite the relative simplicity of their internal structure, we know very little about the internal processes that drive their magnetic dynamos.
Only a limited number of simulations have studied global-scale dynamo processes in M-dwarfs \citep[e.g.,][]{Browning_2008, Yadav_et_al_2015_ApJ, Yadav_et_al_2015_AA, Yadav_et_al_2016}.
Worse, all of these have suffered a fundamental limitation: heretofore it has only been possible to simulate spherical shells, with the coordinate singularity at $r=0$ avoided by making a ``cut-out'' in the middle of the simulation.
The impact of this cut-out on the global solution has been unknown.
Using new techniques in the open-source \dedalus{} pseudospectral framework, we compute the first spherical solutions that include $r=0$ without any cut-out.
These are the first dynamo models of truly, fully convective M-dwarf stars.

As in solar-type stars, our simulations of M-dwarfs have wreath-building convection zone dynamos that create global-scale fields.
However, our M-dwarf simulations find one significant difference:
\textit{We find an abundance of dynamo solutions localized to a single hemisphere}.
Single-hemisphere dynamo states have profound implications for stellar spin-down, exoplanet habitability, and the interpretation of observations of strong fields on stellar surfaces.

\section{Model system}

The interiors of M-dwarfs are regions of low-Mach number convection with significant density stratification.
To study convection and dynamo action in these stars, we solve the anelastic equations using the open-source, spectrally-accurate \dedalus{} framework \citep{Burns_et_al_2020}.
This work is enabled by new approaches to global-spherical domains including $r=0$ \citep{Vasil_et_al_2019, Lecoanet_et_al_2019}.

The non-dimensional energy-conserving anelastic equations with MHD and momentum and thermal diffusion are \citep{Brown_Vasil_Zweibel_2012, Vasil_et_al_2013, Lecoanet_et_al_2014}:
\begin{multline}
\frac{\partial \vec{u}}{\partial t}  + \vec{\del} \pomega  - \frac{\Ra \Ek^2}{ \Pr} T \nabla s - \Ek\frac{ \left(\nabla^2 \vec{u} +\case{1}{3}\del(\del\cdot\vec{u})\right)}{\rho} = \\
- \vec{u}\cdot\vec{\del}\vec{u} - \vec{\hat{z}}\times\vec{u} + \frac{\vec{J}\times \vec{B}}{\rho},
\label{eq:ND momentum}
\end{multline}
\begin{multline}
\frac{\partial s}{\partial t} - \frac{\Ek}{\Pr}\frac{\left(\nabla^2s +\vec{\del} \ln T \cdot \vec{\del} s\right)}{\rho}  =
\\-\vec{u}\cdot\vec{\del}\vec{s}
+ \frac{\Ek}{\Pr}\scrS + \frac{\Pr}{\Ra\Ek}\frac{\Phi_\nu}{\rho T} + \frac{1}{\Pr \Pm}\frac{1}{\Ra\Ek}\frac{J^2}{\rho^2 T},
\label{eq:ND entropy}
\end{multline}
\begin{equation}
\frac{\partial \vec{A}}{\partial t} + \del \phi - \frac{\Ek}{\Pm}\frac{\nabla^2 \vec{A}}{\rho}  = \vec{u}\times \vec{B},
\label{eq:ND induction}
\end{equation}
together with the Coulomb gauge $\vec{\del}\cdot\vec{A} = 0$ and the anelastic $\vec{\del}\cdot\vec{u} + \vec{u} \cdot \vec{\del} \ln \rho = 0$ constraints.
$\scrS$ is a volume heating term and rotation is represented as $\vec{\Omega} = \Omega \vec{\hat{z}}$.
The magnetic field $\vec{B} = \del\times\vec{A}$, and the electrostatic potential $\phi$ and reduced pressure $\pomega$ enforce the constraints.
We use the ideal gas equation of state, and diffusive entropy heat flux $\vec{Q} = -\kappa T \vec{\del} s$ \citep{Lecoanet_et_al_2014}.
The boundaries are constant entropy, impenetrable and stress free for velocity, and potential for magnetic field.
The quantities $\del \ln \rho$, $\del T$, $1/\rho$, $\del \ln T$ and $\scrS$ represent the internal M-dwarf structure.

We non-dimensionalize this set of equations on a characteristic lengthscale $L = R$ the radius of the domain ($0.85 R_\star$), rotation timescale $\tau = 1/(2\Omega)$, density $\rho_c$, magnetic field $B^2 = 16\pi\rho_c \Omega^2 R^2$, and entropy $\scrS_{c}\sigma^{2} R^2/(\kappa/\rho_c)$.
The dynamic momentum, $\mu$, and thermal diffusivities, $\kappa$, are constant. The magnetic diffusivity $\eta \propto \rho^{-1}$.
The global non-dimensional parameters are the Ekman number $\Ek$, Rayleigh number $\Ra$, Prandtl number $\Pr$ and magnetic Prandtl number $\Pm$,
\begin{eqnarray}
 \Ek & = &  \frac{\mu/\rho_c}{2 \Omega R^2}, \quad
   \Ra = \frac{\scrS_{c} \sigma^{2} R^2 T_c}{\mu \kappa^2/\rho_c^3 }, \\
 \quad \Pr & = &  \frac{\mu}{\kappa}, \quad \ \  \quad \, \Pm = \frac{\mu}{\rho_c \eta},
\end{eqnarray}
where $\sigma$ represents the radial extent of the nuclear-burning region. The Rayleigh number has this dependance on the heating because mean entropy contrast across the domain scales with the internal heating to within an $\mathcal{O}(1)$ constant as $\Delta s \sim \rho \, \scrS_{c} \sigma^{2}/\kappa$.
The values are specified at the center $r=0$ (denoted with a subscript $c$), which is the extremal value for $\Ek$ and $\Ra$.
In our simulations, the local Ekman $\Ek$ and Rayleigh $\Ra$ number depend on $\rho(r)/\rho_c$. The Prandtl numbers $\Pr$ and $\Pm$ are constant.

\begin{figure*}[ht!]
  \begin{center}
    \includegraphics{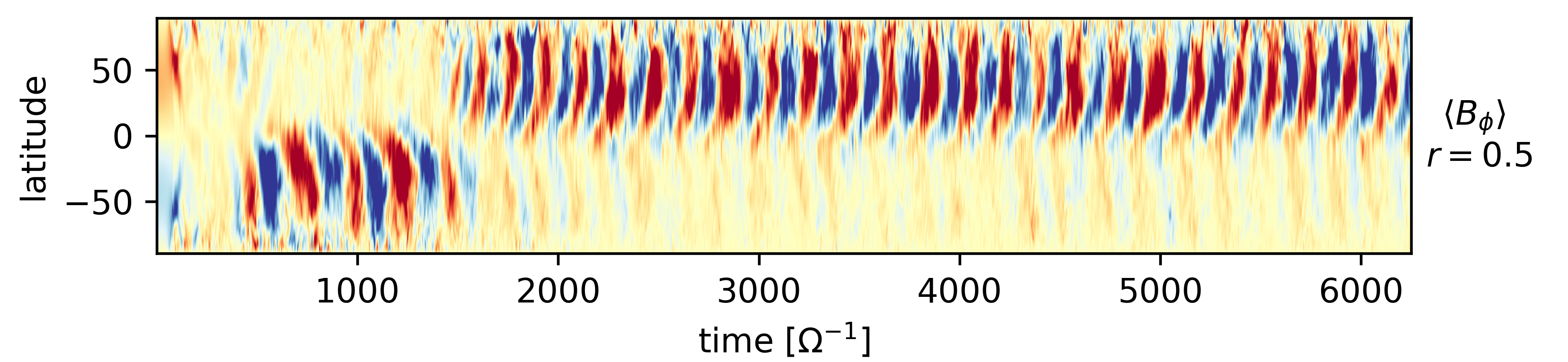}\\[-0.5cm]
    \includegraphics{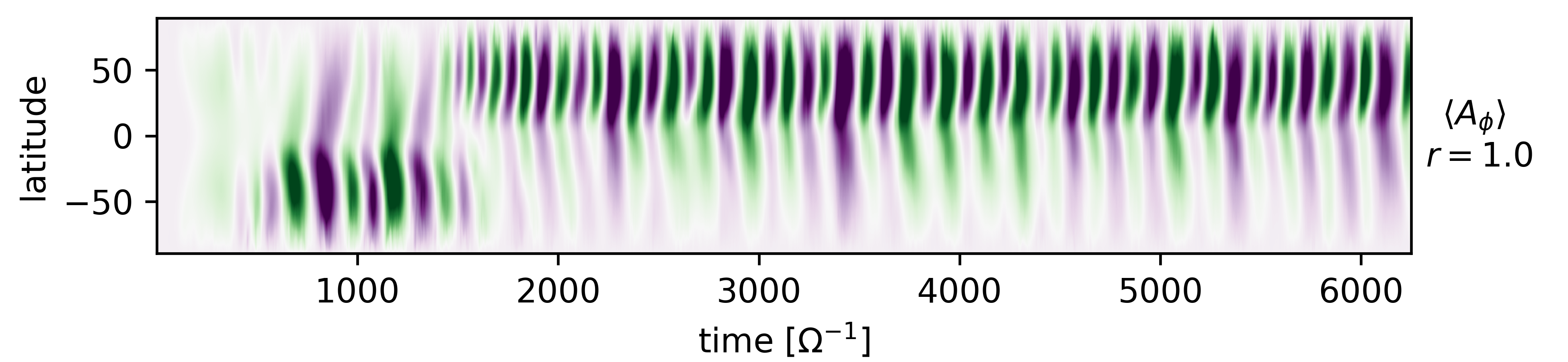}
  \end{center}
  \caption{Time-latitude plots of the global-scale dynamo. Top is the toroidal mangetic field $B_\phi$ at mid-convection zone ($r=0.5$). Bottom is the vector potential of the poloidal magnetic field $A_\phi$ at the top of the domain ($r=1$). Color gives the sense of polarity for both quantities. The system goes through many dynamo reversals of the global-scale field during the period simulated. The mean fields are predominantly in a single hemisphere (here the Northern). At all times the global poloidal fields have a dipolar component, with strong contributions from higher-order modes.
  \label{fig:field timelat}}
\end{figure*}

We use the \textit{convective} Rossby number as an input proxy for the degree of rotational constraint,
\begin{equation}
\RoCsq \equiv \frac{\Ra \Ek^2}{\Pr} = \frac{\scrS_{c} \sigma^{2} T_c }{(\kappa/\rho_c) 4 \Omega^2 R^{2}}.
\end{equation}
This combination has a long history of usefulness for boundary-heated incompressible convection \citep{Julien_et_al_1996}.
In that context, the combination is independent of microphysical diffusivities and depends on the domain size.
The situation here is less simple.
The internal heating structure, $\scrS(r)$, and radiative diffusion $\kappa(r)$, determine the superadiabatic entropy gradient.
Even thought our definition produces a more complex parameter dependance, we find good scaling correlation between $\RoC$ and the output measured local Rossby number; see \citet{Anders_et_al_2019} for more details.

Fully-convective M-dwarf stars are nearly adiabatic, and are well represented by polytropic solutions to the Lane-Emden equations.
We simulate the inner 3 density scale heights ($n_\rho=3$) of an adiabatically stratified ($m=1.5$) Lane-Emden polytrope, corresponding to the inner 85\% of the star.
Our background state is in excellent agreement with a stellar structure model computed with MESA \citep{Paxton_et_al_2011} for a $0.3M_\odot$ star of age 5 Gyrs, at a solar metallicity.\footnote{In contrast to prior work \citep{Yadav_et_al_2015_ApJ, Yadav_et_al_2015_AA, Yadav_et_al_2016} which erroneously assumed a polytropic structure with a gravity profile that grows linearly with radius.}
For $\scrS$, we extract the volumetric heating and radiative cooling from the MESA model, and fit the radial dependence with
\begin{equation}
\scrS = \frac{\epsilon_\mathrm{nuc}}{T} - \frac{1}{\rho T} \del \cdot \vec{F}_\mathrm{rad}
\approx
\left[\left(\frac{Q_0}{Q_1}\right) e^{-r^2/(2 \sigma^2)} + 1\right] Q_1.
\label{eq: NCC source function}
\end{equation}
A nonlinear least squares fit gives $\sigma \approx 0.12$ and $Q_0/Q_1 \approx 11$, with an  $L_2$ error of about 3\%.
The simulations are normalized such that $\scrS(r=0)\sigma^2 = 1$.

In this letter we report on a solution with $\RoC=0.41$, $\Ek=2.5\times10^{-5}$ and $\Pr = \Pm = 1$.
The system starts in an adiabatic state with small entropy noise perturbations.
We initialize a weak poloidal magnetic field with spherical harmonic mode $\ell=1,2$, $m=1$.
The simulation has $L=127$ and $N=127$, where $L$ is the maximal degree of the spin-weighted spherical harmonics used in the spectral representation and $N$ is the maximal degree of the Jacobi polynomials used in the radial expansion.
We computed many such dynamo simulations, but focus our attention on a single representative case.

The convection is a successful dynamo, building both small- and global-scale magnetic fields which change in strength and orientation over time.
This simulation ran for $t=6250\,\Omega^{-1}$, or about 0.3 diffusion times at the core $r=0$ and about 6.3 diffusion times at $r=1$.
The internal heating from $\scrS$ drives a convectively unstable state. By roughly $t\approx 2000\,\Omega^{-1}$, the potential, kinetic and magnetic energies all equilibrate; fluctuating around well-defined averages.

\begin{figure*}[t]
  \begin{center}
  \includegraphics[align=c, width=0.225\textwidth]{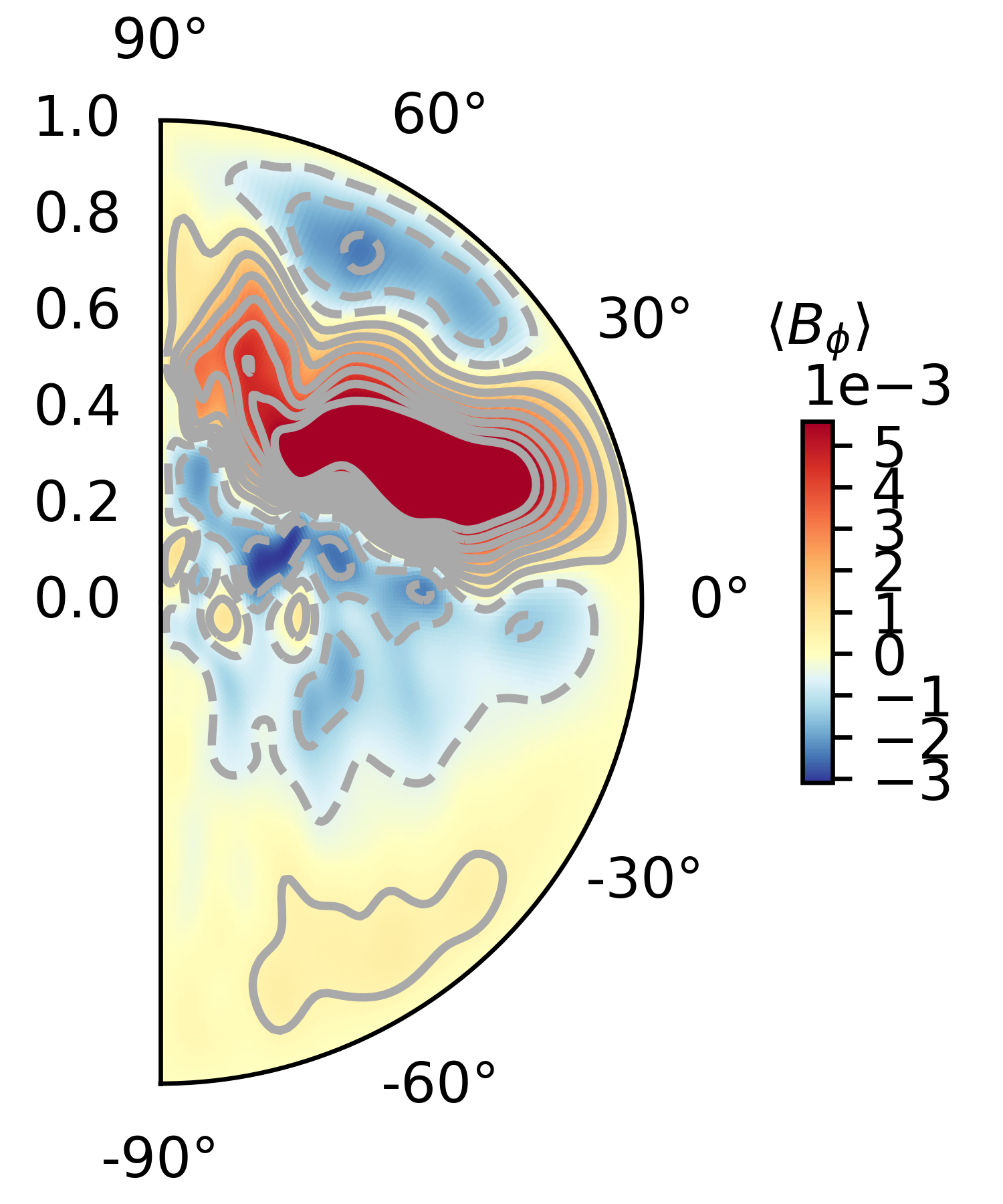}
  \includegraphics[align=c, width=0.225\textwidth]{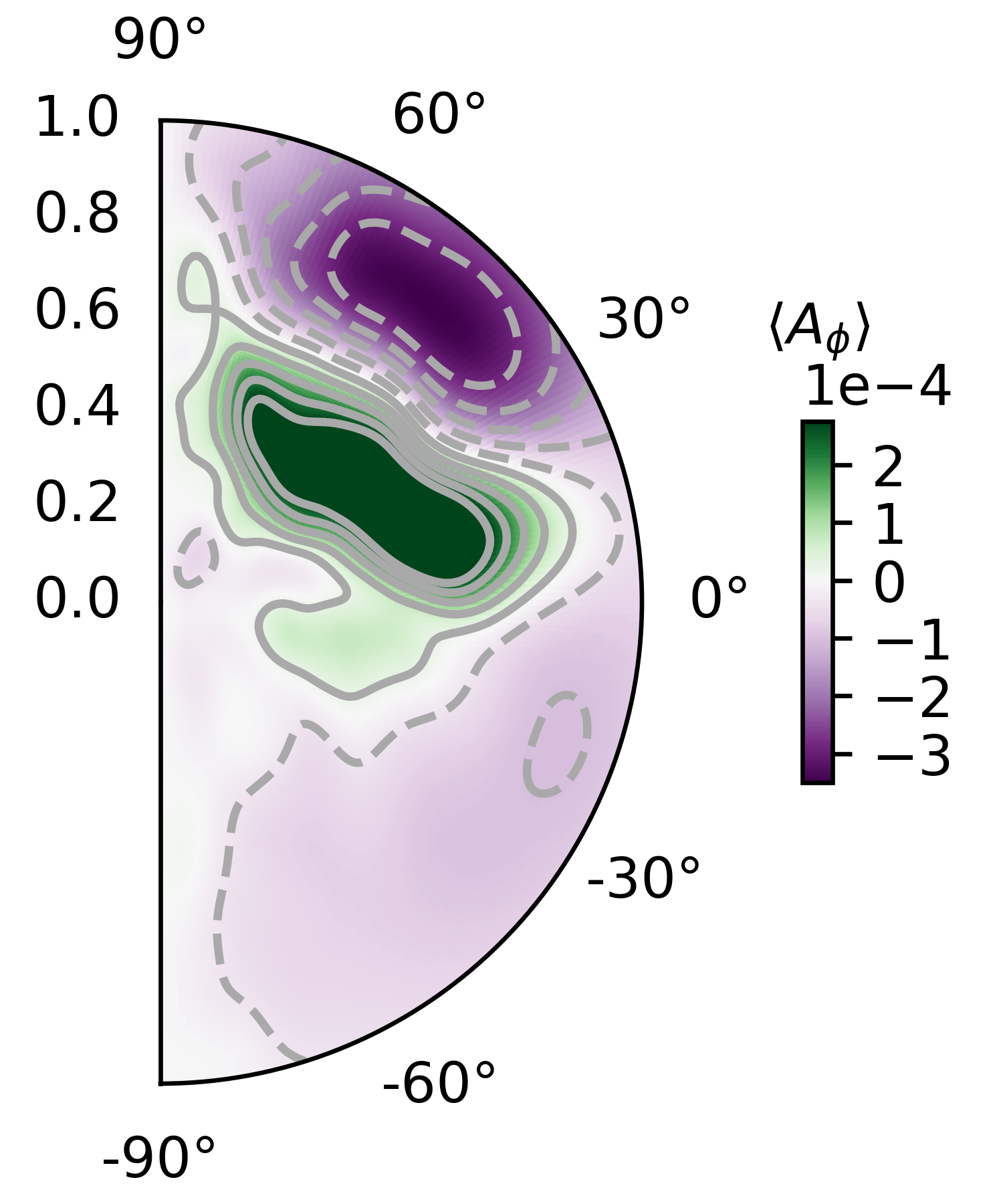}
  \hspace{-0.25cm}\includegraphics[align=c, width=0.5\textwidth]{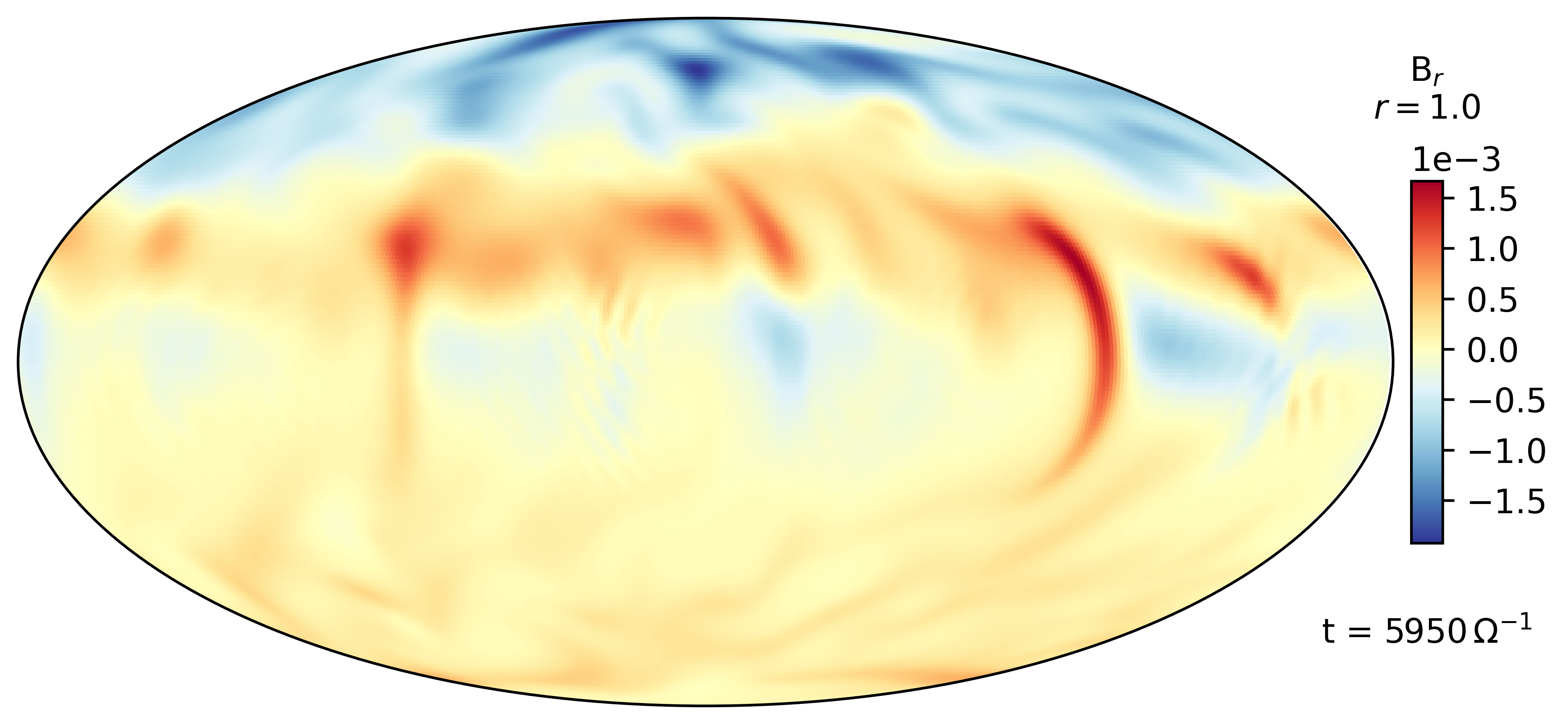}  \\
  \end{center}
  \begin{center}
  \includegraphics[align=c, width=0.225\textwidth]{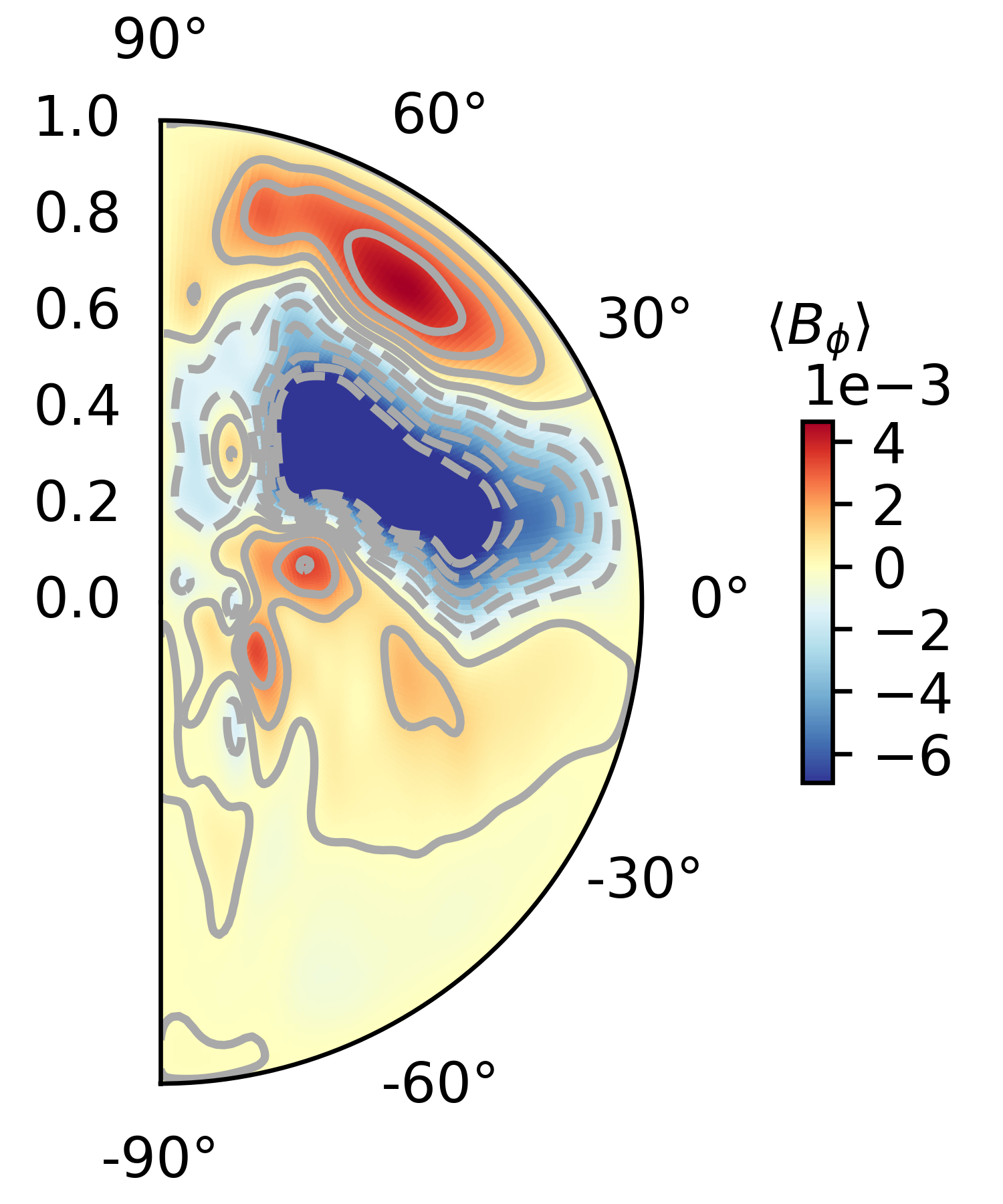}
  \includegraphics[align=c, width=0.225\textwidth]{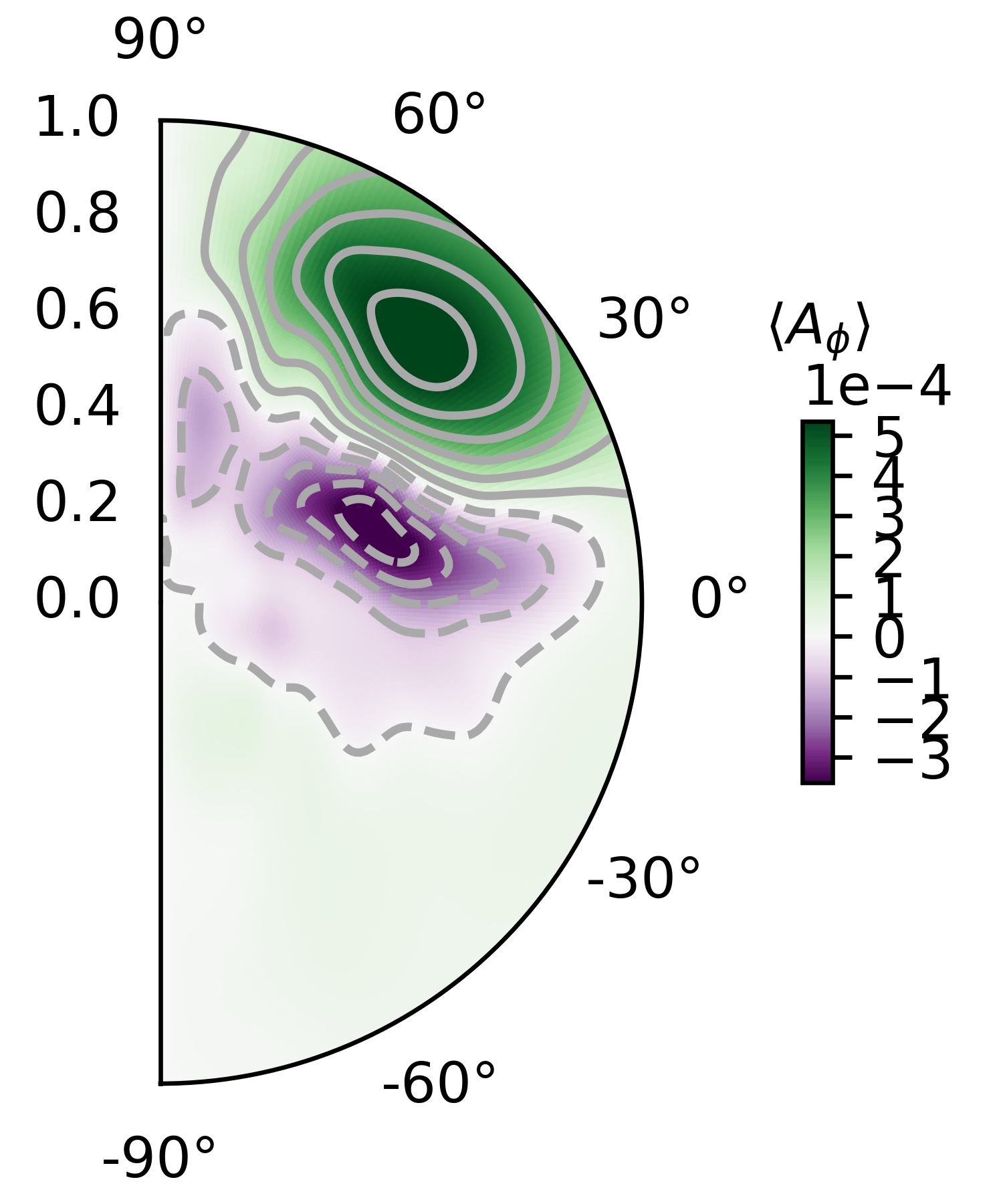}
  \hspace{-0.25cm}\includegraphics[align=c, width=0.5\textwidth]{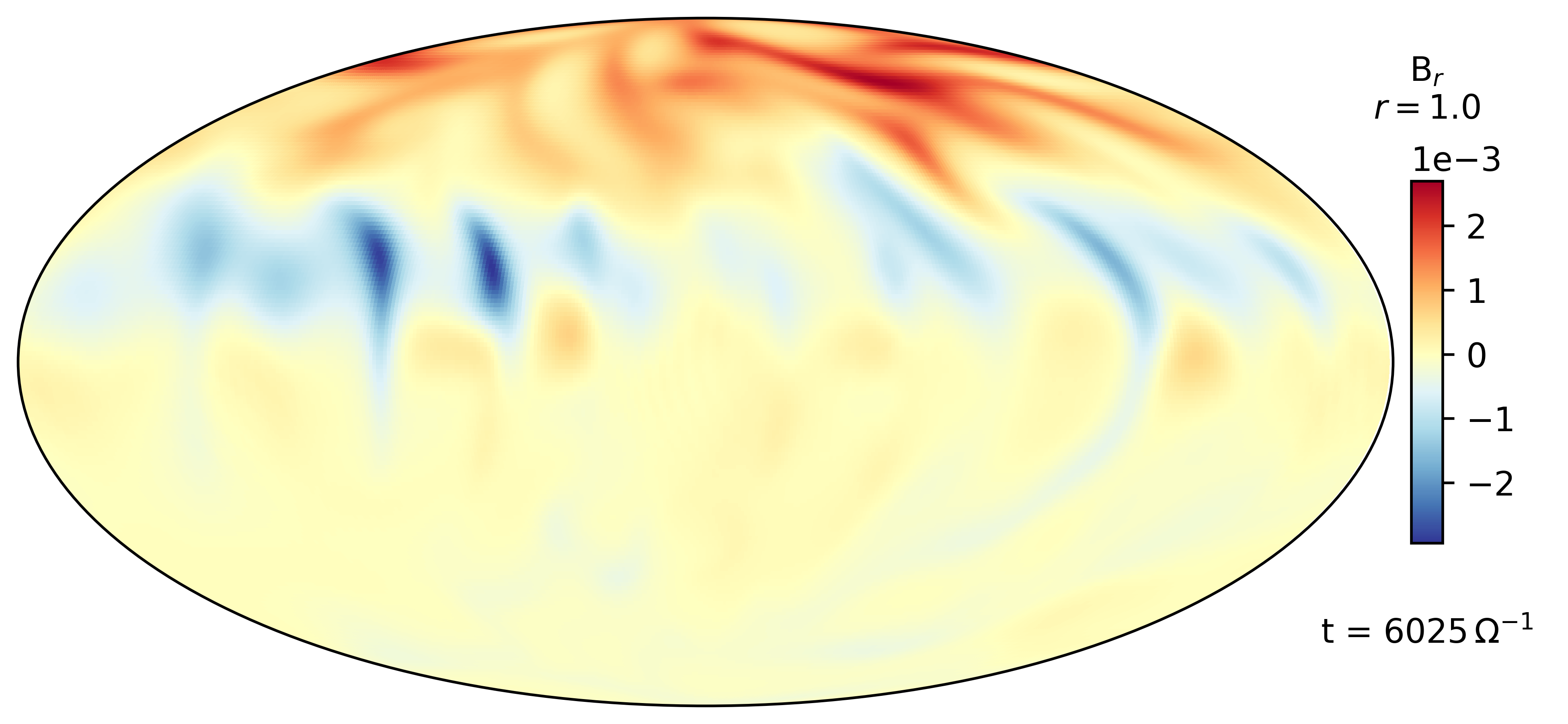}
  \end{center}
  \caption{Magnetic field topologies through a magnetic reversal.  The left side shown the mean toroidal field $B_\phi$ and poloidal vector potential $A_\phi$ for the full domain, temporally averaged over one rotation period. The right side shows the radial magnetic field $B_r$ at the upper boundary $r=1$.  These are taken from periods of peak magnetic field strength during two successive dynamo cycles; the top row from $t=5950\,\Omega^{-1}$, bottom row from $t=6025\,\Omega^{-1}$. \label{fig:fields}}
\end{figure*}

After time-averaging over the second half of the simulation, the RMS Reynolds number $\approx$ 240; the fluctuating Reynolds number $\approx$ 120; the RMS Rossby number $Ro=|\del\times\vec{u}|/2\Omega\approx 0.3$.
In this letter, mean quantities are the $m=0$ components (denoted with angle brackets).
Fluctuating quantities are those with the mean removed.
The time--volume-averaged total magnetic energy is $\approx$ 25\% or the total kinetic energy.
Approximately 25\% of the magnetic energy exists in mean toroidal fields, $\approx$ 68\% exists in fluctuating fields, and the balance is in mean poloidal fields.
The differential rotation contains $\approx$ 45\% of the kinetic energy, with $\approx$ 55\% in the fluctuating convection.

\section{Hemispheric dynamos}

Figure~\ref{fig:field timelat} shows the surprising global-scale mean fields.
The fields are organized in space and in time, and undergo regular periodic reversals of polarity in both the mean toroidal $\langle B_\phi \rangle$ and the mean poloidal field $\langle A_\phi \rangle$.
The fields share similarities with the ``wreathy dynamo'' states found in dynamo simulations of solar-type stars \citep[e.g.,][]{Brown_et_al_2010, Brown_et_al_2011, Nelson_et_al_2013}.
In the fully convective dynamo simulations, the mean fields represent a significant fraction of the total magnetic field.
We measure the global magnetic reversals using Lomb-Scargle periodograms of the mean magnetic field \citep[e.g.,][]{Townsend_2010, VanderPlas_2018}.
The fields change with a period of $P\approx190\,\Omega^{-1}$, which corresponds to a polarity reversal roughly every 15 rotation periods.
The temporal consistency of these cycles is high.
The peak of the periodogram is clearly present in both $\langle B_\phi\rangle$ (normalized peak power $\approx 0.3$, $P\approx 190.1\,\Omega^{-1}$) and $\langle A_\phi \rangle$ (peak power $\approx 0.4$, $P\approx 191.8\,\Omega^{-1}$), while the aliasing peaks are at much lower amplitude ($\lesssim 0.05$ and $\lesssim 0.1$ respectively).
The surface-poloidal and depth-toroidal fields change with nearly the same phase.

It is surprising is that the mean fields of these fully-convective dynamos generally occupy one hemisphere or the other, but not both.
In rapidly rotating solar-type stars with convection-zone shells fields typically symmetrically fill both hemispheres \citep[e.g.,][]{Brown_et_al_2010, Brown_et_al_2011, Nelson_et_al_2013, Augustson_et_al_2013, Augustson_et_al_2015}.
Importantly: \textit{all previous M-dwarf dynamo simulations in deep shells also find two-hemisphere solutions.}
Those simulations have either found mean fields symmetrically in both hemispheres \citep[e.g.,][]{Yadav_et_al_2015_ApJ, Yadav_et_al_2016} or have built mean fields that don't undergo very many reversals \citep{Browning_2008}.

The tendency towards mean fields in a single hemisphere is strong: the initial dynamo action ($t\lesssim 125\,\Omega^{-1}$) is symmetric in the two hemispheres, but strong southern-hemisphere fields rapidly replace the initial transients.
At about $t\approx 1500\,\Omega^{-1}$ those strong fields shift to the northern hemisphere, where they remain for the period we have simulated.
The other hemisphere is not devoid of field, but the mean fields there are much weaker than in the dominant hemisphere.
We speculate that the reversals would continue on some longer timescale if allowed.

Figure~\ref{fig:fields} shows the detailed structure of the global magnetic structure during two phases of reversed polarity late in the simulation.
The first happens around $t\approx5950\,\Omega^{-1}$ averaged over one rotation period ($\Delta t \approx 2 \pi \,\Omega^{-1}$) when positive (red) polarity dominates the northern hemisphere.
We see that a positive $\langle B_\phi\rangle$ wreath fills the star, reaching from the core to the surface.
North of this structure and closer to the surface lie the weak remnant wreaths of the previous cycle.
A distinctly weaker and non-dipolar poloidal field $\langle A_\phi\rangle$ accompanies the dominant toroidal structure.
There is some global-dipole component, but it is much weaker than the higher-order modes.
The near-surface cuts also show the poloidal-field structure; indicating a substantial quadrupolar component.
As the wreath migrates up along the rotation axis, a new opposite-polarity wreath (blue) replaces it.
This reversal is visible in the surface radial fields.

\begin{figure*}
  \begin{center}
    \includegraphics[align=c, width=0.35\linewidth]{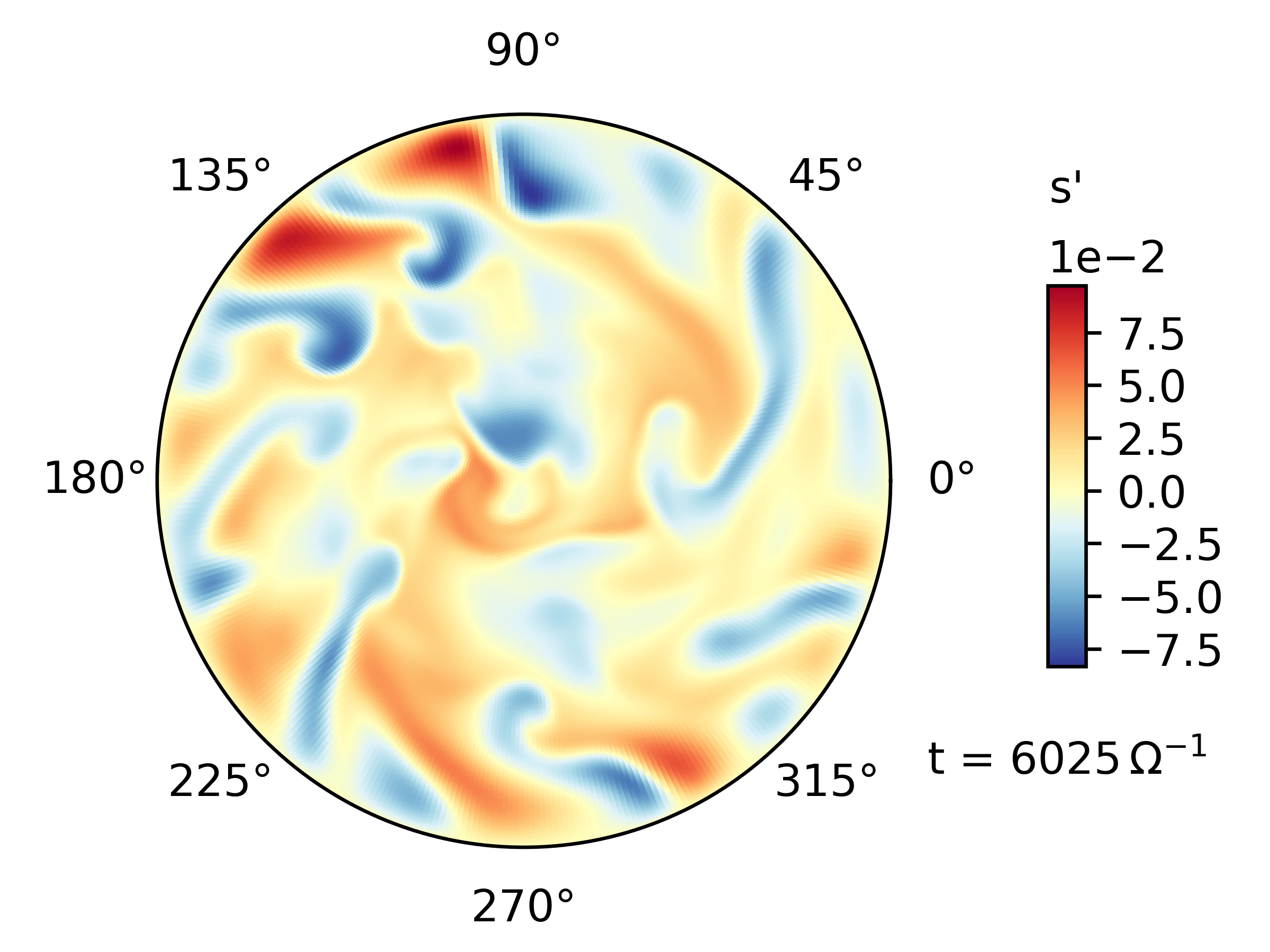}
    \includegraphics[align=c, width=0.4\linewidth]{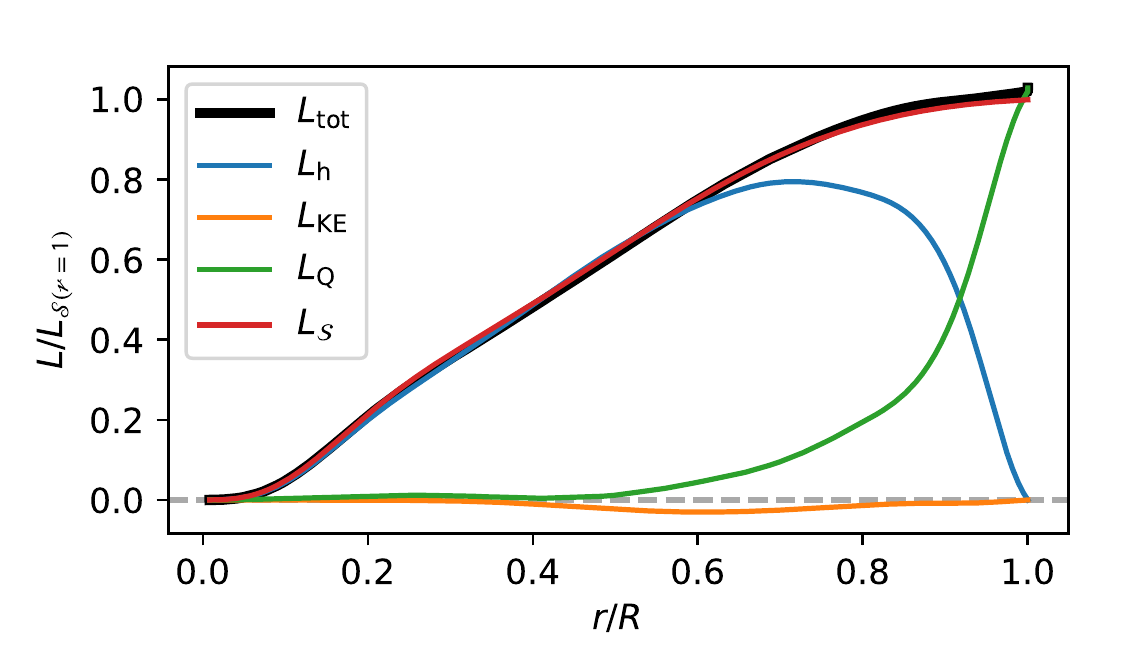}
    \includegraphics[align=c, width=0.225\linewidth]{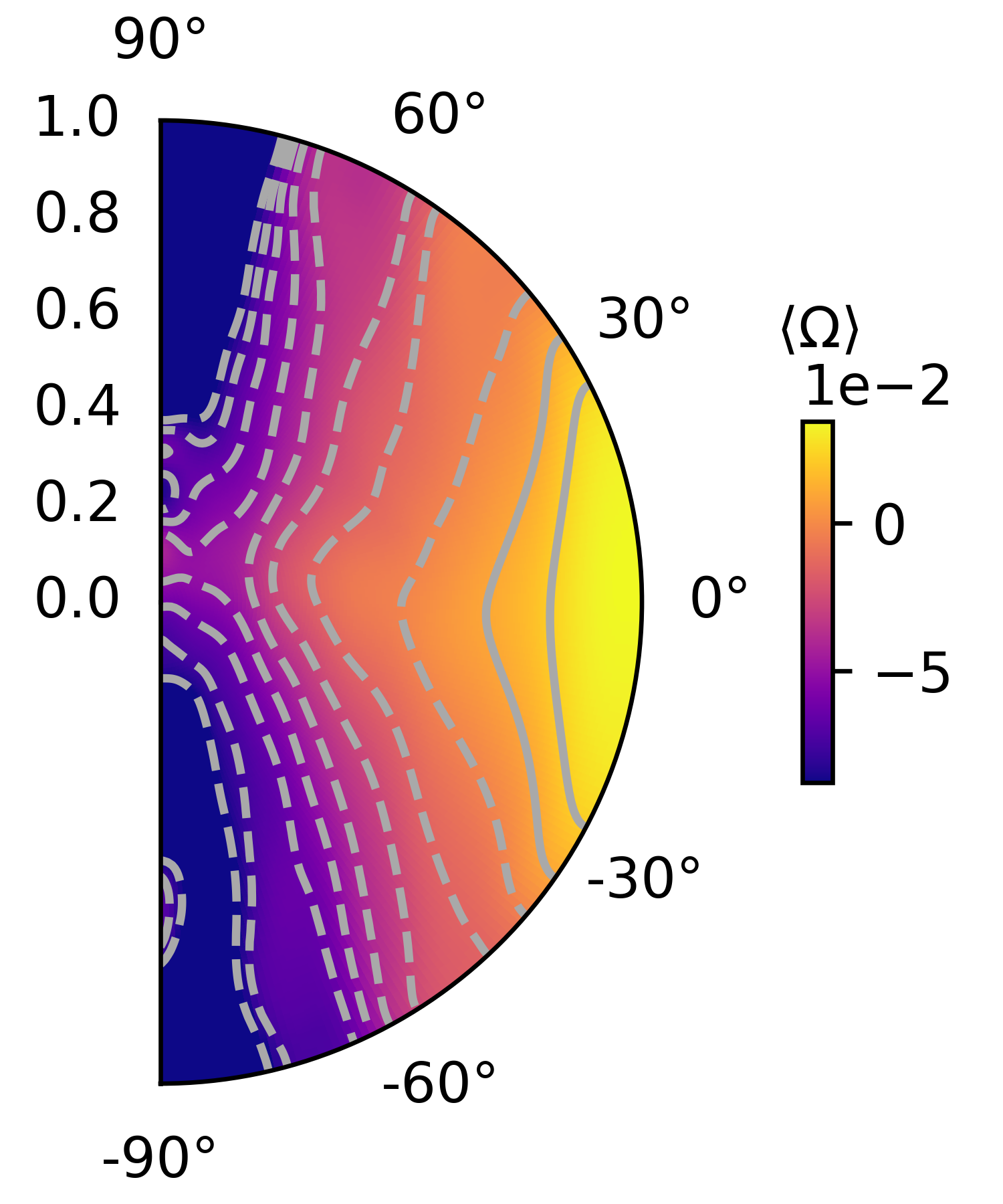}
  \end{center}
  \caption{Properties of flows and heat transport in the fully convective simulation. Left: a cut across the equator of the fluctuating entropy with the $m=0$ average subtracted.  Center: Luminosity profiles throughout the simulation.  The radial profiles are averaged horizontally and in time.  Right: These flows drive a strong solar-like differential rotation; fast near the equator and slow near the rotation axis.  These flows are shown at time $t=6025\,\Omega^{-1}$, corresponding to the lower row of Figure~\ref{fig:fields}; the differential rotation is time-averaged over about 10 rotation periods, centered on this time.
  \label{fig:entropy}}
\end{figure*}

The convective flows fill the interior, as visible in Figure~\ref{fig:entropy}.
Here we show entropy fluctuations in a cut at the equator.
Cold, low-entropy plumes (blue) fall from the near-surface regions.
The core comprises rising warm, high-entropy plumes (red).
The center at $r=0$ is a region of vigorous dynamics, and plumes frequently cross the origin.
Convective enthalpy transport ($L_h$) dominates the radial luminosity balance, except in the upper boundary layer where thermal diffusion takes over ($L_Q$).
Though these simulations are stratified, the kinetic energy flux ($L_\mathrm{KE}$) is small, similar to rotationally-constrained simulations of solar-type stars \citep[e.g.,][]{Brown_et_al_2008}.
The sum of the luminosities closely matches the nuclear source derived from MESA ($L_\scrS$) indicating that these models are in proper thermal equilibrium.
This convection drives a robust differential rotation, shown by the mean angular velocity, $\langle \Omega \rangle$, with a fast equator and slow polar and core regions.
The gradients of angular velocity are strong in both radius and latitude. These gradients likely build the mean toroidal field $\langle B_\phi \rangle$ via shear induction (a.k.a the "$\Omega$-effect").
Correlations in the fluctuating flows and fields maintain the polar field.
The latitudinal contrast in $\langle \Omega \rangle$ is about 10\%, and the mean magnetic fields coexist with this shearing flow.
This is similar to results from rapidly rotating simulations of solar-type stars \citep{Brown_et_al_2011}, but markedly different from the results of \citet{Browning_2008} or \cite{Yadav_et_al_2015_ApJ} where the fields strongly quench the differential rotation.
Whether differential rotation and fields can coexist likely depends on the Rossby number regimes of the dynamos, as discussed in \cite{Yadav_et_al_2016}.

\section{Implications of Hemispheric Dynamos}

This letter has focused on a single dynamo solution that shows striking hemispheric asymmetries.
We find the same preference for single-hemisphere dynamos in every fully convective M-dwarf case computed to present; including a sweep of cases at $\Ek=2.5\times10^{-5}$ and $\RoC = 0.38$--$0.57$, corresponding to RMS Reynolds numbers of $210$--$350$ and Rossby numbers of $0.27$--$0.42$.
To check if single-hemisphere preference happens spuriously from a numerical source, we also rotated around the $\vec{\hat{x}}$ rather than the $\vec{\hat{z}}$ axis (for $\RoC = 0.41$, $\Ek=2.5\times10^{-5}$).
In this strange solution, cycling single-hemispheric dynamo action happened around the east-west pole, relative to the numerical grid.

So far, we cannot determine why fully convective simulations so strongly prefer single-hemisphere dynamos; nor why these have not been found previously.
Possibly, hemispheric dynamos are local to this region of parameter space.
It is also possible that small core cutouts, as used in shell simulations, significantly change the allowed topologies of the global-scale fields even when those cutouts are small ($\lesssim 10\% R$).
Small solid objects are a relevant source of vorticity injection.
In any case, we find that hemispheric dynamo states are robust and possibly ubiquitous in fully convective simulations, and this prompts us to consider their impacts if they exist in real stars.

Strong global-scale mean fields confined to a single hemisphere will couple dramatically differently to the outflowing stellar wind than a traditional dipole.
The higher-order moments of field fall of much faster with distance and this likely produces weaker angular momentum transport and less efficient spindown in hemispheric-dynamo stars.
The presence of mean-field in a single hemisphere could dramatically affect attempts to observe this field and lead to significant viewing angle effects; this could explain substantial differences in observed topologies for otherwise similar stars \citep[e.g.,][]{Morin_et_al_2010}.

Lastly, we expect that the habitability around a hemispheric-dynamo star might significantly differ from one with a dipolar field.
Hemispheric-dynamo stars would be less likely to turn nearby exoplanets into lava worlds via inductive heating \citep[e.g.,][]{Kislyakova_et_al_2017, Kislyakova_et_al_2018}, as the stellar field at the location of the planetary orbit will be much weaker.
But the bad news is there might be much higher levels of magnetic activity, flares and coronal mass ejections; the mean-field topologies are more complicated than simple dipoles.
Alternatively, though, space weather might show a strong directional preference.
A planet with a low-inclination orbit might escape the worst.
But even then we anticipate the preferred hemisphere to switch often enough to cause trouble for everyone.
Altogether, fully convective stars have a rich landscape of dynamo states, and exploring this landscape shines new light on our understanding of the dynamos operating in solar-type stars. It seems many questions remain wide open.

\acknowledgements
This work was supported by NASA LWS grant NNX16AC92G, NASA SSW grant 80NSSC19K0026, and NASA HTMS grant 80NSSC20K1280.  Computations were conducted with support by the NASA High End Computing (HEC) Program through the NASA Advanced Supercomputing (NAS) Division at Ames Research Center on Pleiades with allocation GIDs s1647 and s2114.

\software{
Dedalus \citep[commit \texttt{7efb884},][\url{http://dedalus-project.org}]{Burns_et_al_2020},
Dedalus-Sphere \citep[commit \texttt{c663ff4},][\url{https://github.com/DedalusProject/dedalus_sphere}]{Vasil_et_al_2019, Lecoanet_et_al_2019},
MESA \citep[release \texttt{11701},][\url{http://mesa.sourceforge.net/}]{Paxton_et_al_2011, Paxton_et_al_2013, Paxton_et_al_2015, Paxton_et_al_2018}.
}

\bibliographystyle{aasjournal}
\bibliography{mdwarf}

\begin{thebibliography}{}
\expandafter\ifx\csname natexlab\endcsname\relax\def\natexlab#1{#1}\fi
\providecommand{\url}[1]{\href{#1}{#1}}
\providecommand{\dodoi}[1]{doi:~\href{http://doi.org/#1}{\nolinkurl{#1}}}
\providecommand{\doeprint}[1]{\href{http://ascl.net/#1}{\nolinkurl{http://ascl.net/#1}}}
\providecommand{\doarXiv}[1]{\href{https://arxiv.org/abs/#1}{\nolinkurl{https://arxiv.org/abs/#1}}}

\bibitem[{{Anders} {et~al.}(2019){Anders}, {Manduca}, {Brown}, {Oishi}, \&
  {Vasil}}]{Anders_et_al_2019}
{Anders}, E.~H., {Manduca}, C.~M., {Brown}, B.~P., {Oishi}, J.~S., \& {Vasil},
  G.~M. 2019, \apj, 872, 138, \dodoi{10.3847/1538-4357/aaff61}

\bibitem[{{Augustson} {et~al.}(2015){Augustson}, {Brun}, {Miesch}, \&
  {Toomre}}]{Augustson_et_al_2015}
{Augustson}, K., {Brun}, A.~S., {Miesch}, M., \& {Toomre}, J. 2015, \apj, 809,
  149, \dodoi{10.1088/0004-637X/809/2/149}

\bibitem[{{Augustson} {et~al.}(2013){Augustson}, {Brun}, \&
  {Toomre}}]{Augustson_et_al_2013}
{Augustson}, K.~C., {Brun}, A.~S., \& {Toomre}, J. 2013, \apj, 777, 153,
  \dodoi{10.1088/0004-637X/777/2/153}

\bibitem[{{Brown} {et~al.}(2008){Brown}, {Browning}, {Brun}, {Miesch}, \&
  {Toomre}}]{Brown_et_al_2008}
{Brown}, B.~P., {Browning}, M.~K., {Brun}, A.~S., {Miesch}, M.~S., \& {Toomre},
  J. 2008, \apj, 689, 1354, \dodoi{10.1086/592397}

\bibitem[{{Brown} {et~al.}(2010){Brown}, {Browning}, {Brun}, {Miesch}, \&
  {Toomre}}]{Brown_et_al_2010}
---. 2010, \apj, 711, 424, \dodoi{10.1088/0004-637X/711/1/424}

\bibitem[{{Brown} {et~al.}(2011){Brown}, {Miesch}, {Browning}, {Brun}, \&
  {Toomre}}]{Brown_et_al_2011}
{Brown}, B.~P., {Miesch}, M.~S., {Browning}, M.~K., {Brun}, A.~S., \& {Toomre},
  J. 2011, \apj, 731, 69, \dodoi{10.1088/0004-637X/731/1/69}

\bibitem[{{Brown} {et~al.}(2012){Brown}, {Vasil}, \&
  {Zweibel}}]{Brown_Vasil_Zweibel_2012}
{Brown}, B.~P., {Vasil}, G.~M., \& {Zweibel}, E.~G. 2012, \apj, 756, 109,
  \dodoi{10.1088/0004-637X/756/2/109}

\bibitem[{{Browning}(2008)}]{Browning_2008}
{Browning}, M.~K. 2008, \apj, 676, 1262, \dodoi{10.1086/527432}

\bibitem[{{Burns} {et~al.}(2020){Burns}, {Vasil}, {Oishi}, {Lecoanet}, \&
  {Brown}}]{Burns_et_al_2020}
{Burns}, K.~J., {Vasil}, G.~M., {Oishi}, J.~S., {Lecoanet}, D., \& {Brown},
  B.~P. 2020, Physical Review Research, 2, 023068,
  \dodoi{10.1103/PhysRevResearch.2.023068}

\bibitem[{{Dressing} \& {Charbonneau}(2015)}]{Dressing_Charbonneau_2015}
{Dressing}, C.~D., \& {Charbonneau}, D. 2015, \apj, 807, 45,
  \dodoi{10.1088/0004-637X/807/1/45}

\bibitem[{{Jackson} \& {Jeffries}(2012)}]{Jackson_Jeffries_2012}
{Jackson}, R.~J., \& {Jeffries}, R.~D. 2012, \mnras, 423, 2966,
  \dodoi{10.1111/j.1365-2966.2012.21119.x}

\bibitem[{{Julien} {et~al.}(1996){Julien}, {Legg}, {McWilliams}, \&
  {Werne}}]{Julien_et_al_1996}
{Julien}, K., {Legg}, S., {McWilliams}, J., \& {Werne}, J. 1996, Journal of
  Fluid Mechanics, 322, 243, \dodoi{10.1017/S0022112096002789}

\bibitem[{{Kislyakova} {et~al.}(2018){Kislyakova}, {Fossati}, {Johnstone},
  {Noack}, {L{\"u}ftinger}, {Zaitsev}, \& {Lammer}}]{Kislyakova_et_al_2018}
{Kislyakova}, K.~G., {Fossati}, L., {Johnstone}, C.~P., {et~al.} 2018, \apj,
  858, 105, \dodoi{10.3847/1538-4357/aabae4}

\bibitem[{{Kislyakova} {et~al.}(2017){Kislyakova}, {Noack}, {Johnstone},
  {Zaitsev}, {Fossati}, {Lammer}, {Khodachenko}, {Odert}, \&
  {G{\"u}del}}]{Kislyakova_et_al_2017}
{Kislyakova}, K.~G., {Noack}, L., {Johnstone}, C.~P., {et~al.} 2017, Nature
  Astronomy, 1, 878, \dodoi{10.1038/s41550-017-0284-0}

\bibitem[{{Lecoanet} {et~al.}(2014){Lecoanet}, {Brown}, {Zweibel}, {Burns},
  {Oishi}, \& {Vasil}}]{Lecoanet_et_al_2014}
{Lecoanet}, D., {Brown}, B.~P., {Zweibel}, E.~G., {et~al.} 2014, \apj, 797, 94,
  \dodoi{10.1088/0004-637X/797/2/94}

\bibitem[{Lecoanet {et~al.}(2019)Lecoanet, Vasil, Burns, Brown, \&
  Oishi}]{Lecoanet_et_al_2019}
Lecoanet, D., Vasil, G.~M., Burns, K.~J., Brown, B.~P., \& Oishi, J.~S. 2019,
  Journal of Computational Physics: X, 100012

\bibitem[{{Morin} {et~al.}(2010){Morin}, {Donati}, {Petit}, {Delfosse},
  {Forveille}, \& {Jardine}}]{Morin_et_al_2010}
{Morin}, J., {Donati}, J.~F., {Petit}, P., {et~al.} 2010, \mnras, 407, 2269,
  \dodoi{10.1111/j.1365-2966.2010.17101.x}

\bibitem[{{Nelson} {et~al.}(2013){Nelson}, {Brown}, {Brun}, {Miesch}, \&
  {Toomre}}]{Nelson_et_al_2013}
{Nelson}, N.~J., {Brown}, B.~P., {Brun}, A.~S., {Miesch}, M.~S., \& {Toomre},
  J. 2013, \apj, 762, 73, \dodoi{10.1088/0004-637X/762/2/73}

\bibitem[{{Newton} {et~al.}(2016){Newton}, {Irwin}, {Charbonneau},
  {Berta-Thompson}, {Dittmann}, \& {West}}]{Newton_et_al_2016}
{Newton}, E.~R., {Irwin}, J., {Charbonneau}, D., {et~al.} 2016, \apj, 821, 93,
  \dodoi{10.3847/0004-637X/821/2/93}

\bibitem[{{Newton} {et~al.}(2018){Newton}, {Mondrik}, {Irwin}, {Winters}, \&
  {Charbonneau}}]{Newton_et_al_2018}
{Newton}, E.~R., {Mondrik}, N., {Irwin}, J., {Winters}, J.~G., \&
  {Charbonneau}, D. 2018, \aj, 156, 217, \dodoi{10.3847/1538-3881/aad73b}

\bibitem[{{Paxton} {et~al.}(2011){Paxton}, {Bildsten}, {Dotter}, {Herwig},
  {Lesaffre}, \& {Timmes}}]{Paxton_et_al_2011}
{Paxton}, B., {Bildsten}, L., {Dotter}, A., {et~al.} 2011, \apjs, 192, 3,
  \dodoi{10.1088/0067-0049/192/1/3}

\bibitem[{{Paxton} {et~al.}(2013){Paxton}, {Cantiello}, {Arras}, {Bildsten},
  {Brown}, {Dotter}, {Mankovich}, {Montgomery}, {Stello}, \&
  {Timmes}}]{Paxton_et_al_2013}
{Paxton}, B., {Cantiello}, M., {Arras}, P., {et~al.} 2013, \apjs, 208, 4,
  \dodoi{10.1088/0067-0049/208/1/4}

\bibitem[{{Paxton} {et~al.}(2015){Paxton}, {Marchant}, {Schwab}, {Bauer},
  {Bildsten}, {Cantiello}, {Dessart}, {Farmer}, {Hu}, \&
  {Langer}}]{Paxton_et_al_2015}
{Paxton}, B., {Marchant}, P., {Schwab}, J., {et~al.} 2015, \apjs, 220, 15,
  \dodoi{10.1088/0067-0049/220/1/15}

\bibitem[{{Paxton} {et~al.}(2018){Paxton}, {Schwab}, {Bauer}, {Bildsten},
  {Blinnikov}, {Duffell}, {Farmer}, {Goldberg}, {Marchant}, \&
  {Sorokina}}]{Paxton_et_al_2018}
{Paxton}, B., {Schwab}, J., {Bauer}, E.~B., {et~al.} 2018, \apjs, 234, 34,
  \dodoi{10.3847/1538-4365/aaa5a8}

\bibitem[{{Schmidt} {et~al.}(2019){Schmidt}, {Shappee}, {van Saders}, {Stanek},
  {Brown}, {Kochanek}, {Dong}, {Drout}, {Frank}, {Holoien}, {Johnson},
  {Madore}, {Prieto}, {Seibert}, {Seidel}, \& {Simonian}}]{Schmidt_et_al_2019}
{Schmidt}, S.~J., {Shappee}, B.~J., {van Saders}, J.~L., {et~al.} 2019, \apj,
  876, 115, \dodoi{10.3847/1538-4357/ab148d}

\bibitem[{{Shkolnik} \& {Barman}(2014)}]{Shkolnik_et_al_2014}
{Shkolnik}, E.~L., \& {Barman}, T.~S. 2014, \aj, 148, 64,
  \dodoi{10.1088/0004-6256/148/4/64}

\bibitem[{{Townsend}(2010)}]{Townsend_2010}
{Townsend}, R.~H.~D. 2010, \apjs, 191, 247, \dodoi{10.1088/0067-0049/191/2/247}

\bibitem[{{VanderPlas}(2018)}]{VanderPlas_2018}
{VanderPlas}, J.~T. 2018, \apjs, 236, 16, \dodoi{10.3847/1538-4365/aab766}

\bibitem[{{Vasil} {et~al.}(2013){Vasil}, {Lecoanet}, {Brown}, {Wood}, \&
  {Zweibel}}]{Vasil_et_al_2013}
{Vasil}, G.~M., {Lecoanet}, D., {Brown}, B.~P., {Wood}, T.~S., \& {Zweibel},
  E.~G. 2013, \apj, 773, 169, \dodoi{10.1088/0004-637X/773/2/169}

\bibitem[{Vasil {et~al.}(2019)Vasil, Lecoanet, Burns, Oishi, \&
  Brown}]{Vasil_et_al_2019}
Vasil, G.~M., Lecoanet, D., Burns, K.~J., Oishi, J.~S., \& Brown, B.~P. 2019,
  Journal of Computational Physics: X, 100013

\bibitem[{{West} {et~al.}(2015){West}, {Weisenburger}, {Irwin},
  {Berta-Thompson}, {Charbonneau}, {Dittmann}, \& {Pineda}}]{West_et_al_2015}
{West}, A.~A., {Weisenburger}, K.~L., {Irwin}, J., {et~al.} 2015, \apj, 812, 3,
  \dodoi{10.1088/0004-637X/812/1/3}

\bibitem[{{Yadav} {et~al.}(2015{\natexlab{a}}){Yadav}, {Christensen}, {Morin},
  {Gastine}, {Reiners}, {Poppenhaeger}, \& {Wolk}}]{Yadav_et_al_2015_ApJ}
{Yadav}, R.~K., {Christensen}, U.~R., {Morin}, J., {et~al.} 2015{\natexlab{a}},
  \apjl, 813, L31, \dodoi{10.1088/2041-8205/813/2/L31}

\bibitem[{{Yadav} {et~al.}(2016){Yadav}, {Christensen}, {Wolk}, \&
  {Poppenhaeger}}]{Yadav_et_al_2016}
{Yadav}, R.~K., {Christensen}, U.~R., {Wolk}, S.~J., \& {Poppenhaeger}, K.
  2016, \apjl, 833, L28, \dodoi{10.3847/2041-8213/833/2/L28}

\bibitem[{{Yadav} {et~al.}(2015{\natexlab{b}}){Yadav}, {Gastine},
  {Christensen}, \& {Reiners}}]{Yadav_et_al_2015_AA}
{Yadav}, R.~K., {Gastine}, T., {Christensen}, U.~R., \& {Reiners}, A.
  2015{\natexlab{b}}, \aap, 573, A68, \dodoi{10.1051/0004-6361/201424589}

\end{thebibliography}

\end{document}